\documentclass[conference]{IEEEtran}
\IEEEoverridecommandlockouts
\renewcommand\IEEEkeywordsname{Keywords}
\usepackage{cite}
\usepackage{subfloat}
\usepackage{amsmath,amssymb,amsfonts}
\usepackage{algorithmic}
\usepackage{graphicx}
\usepackage{textcomp}
\usepackage{xcolor}
\usepackage{wrapfig}
\usepackage{color,soul}
\usepackage{multirow}
\usepackage{url}
\usepackage{tikz}
\usepackage{comment}
\usepackage{subcaption}
\usepackage{graphicx}
\usepackage[absolute]{textpos}
\usepackage[flushleft]{threeparttable}
\def\checkmark{\tikz\fill[scale=0.4](0,.35) -- (.25,0) -- (1,.7) -- (.25,.15) -- cycle;} 

\usepackage{atbegshi}
\AtBeginDocument{\AtBeginShipoutNext{\AtBeginShipoutDiscard}}

\begin{document}

\title{Study on Patterns and Effect of Task Diversity\\ in Software Crowdsourcing}

\author{
Denisse Martinez Mejorado, Razieh Saremi, Ye Yang, and Jose E. Ramirez-Marquez\\
Stevens Institute of Technology, School of Systems and Enterprises\\
Hoboken, NJ, United States \\
\{dmartin2,rsaremi,yyang4,jmarquez\}@stevens.edu
}

\thanks{Permission to make digital or hard copies of all or part of this work for personal or classroom use is granted without fee provided that copies are not made or distributed for profit or commercial advantage and that copies bear this notice and the full citation on the first page. Copyrights for components of this work owned by others than the author(s) must be honored. Abstracting with credit is permitted. To copy otherwise, or republish, to post on servers or to redistribute to lists, requires prior specific permission and/or a fee. Request permissions from Permissions@acm.org.\\ ESEM '20, October 8–9, 2020, Bari, Italy\\
© 2020 Copyright is held by the owner/author(s). Publication rights licensed to ACM. ACM ISBN 978-1-4503-7580-1/20/10…\$15.00 \\
https://doi.org/10.1145/3382494.3410689}

\maketitle

\begin{abstract}
Context: The success of software crowdsourcing depends on steady pools of task demand and active workers supply. Existing analysis reveals an average task failure ratio of 15.7\% in software crowdsourcing market.

Goal: The objective of this study is to empirically investigate patterns and effect of task diversity in software crowdsourcing platform in order to improve the success and efficiency of software crowdsourcing.

Method: We first propose a conceptual task diversity model, and develop an approach to measuring and analyzing task diversity. More specifically, task diversity is characterized based on semantic similarity, dynamic competition level, and the analysis includes identifying the dominant attributes distinguishing the competition levels, and measuring the impact of task diversity on task success and worker performance in crowdsourcing platform. The empirical study is conducted on more than one year’s real-world data from TopCoder, one of the leading software crowdsourcing platforms.

Results: We identified that monetary prize and task complexity are the dominant attributes that differentiate among different competition levels. Based on these dominant attributes, we concluded three task diversity patterns (configurations) from workers behavior perspective: responsive-to-prize, responsive-to-prize-and-complexity and over-responsive-to-prize. This study supports that the second pattern, i.e. responsive-to-prize-and-complexity configuration, associates with the lowest task failure ratio.

Conclusions: These findings are helpful for task requesters to plan for and improve task success in a more effective and efficient manner in software crowdsourcing platform.

\end{abstract}




\section{Introduction}

Crowdsourced software development (CSD) has gained increased popularity in recent years for its extraordinary benefits. Examples of such benefits include cost reduction, faster time-to-market, higher quality, creativity and open innovation \cite{stol2014two}, attributed to its direct access to infinite, diverse talent pool \cite{lakhani2010topcoder, zanatta2016software}.

Generally, decisions regarding to crowdsourcing a software task usually involves organizational and technical factors such as characteristics of the software problem, knowledge and expertise required for deriving viable solutions, and the size of the crowd pool \cite{thuan2016factors}, etc. Due to the lack of transparency between task requesters and crowd workers, this emergent CSD paradigm has introduced inherent risks in planning and managing crowdsourced projects \cite{stol2014researching} \cite{lakhani2007value} \cite{lakhani2010topcoder} \cite{yu2015efficient}.

From a task requester’s perspective, top risks associated with crowdsourced tasks include uncertainty on both the number and trustworthiness of registrants and quality of the received submissions from the unknown workers \cite{yu2015efficient} \cite{yang2015award} \cite{eickhoff2013increasing}. Existing studies reveal a 82.9\% task-dropping rate in software crowdsourcing marketplace, which leads to a 15.7\% of crowdsourcing failure \cite {yang2016should}.

From crowd workers' perspective, they usually choose to register, work and submit for tasks based on some personal utility algorithm, their skills and some unknown factors \cite{faradani2011s}. It is also reported that they prefer to continue working on similar context tasks in terms of task type, required technology and platform, and their previous experience \cite{difallah2016scheduling}\cite{crump2013evaluating} \cite{yin2014monetary}. 

Therefore, understanding, attracting, and improving worker engagement is one of the key challenges in successful software crowdsourcing work \cite{kittur2013future}. For example, higher density of tasks, i.e. a large number of similar tasks during a given time period, in a crowdsourcing market may offer different task opportunities to crowd workers and increase task competition intensity, and consequently lead to higher chance of task starvation or cancellation.

It is essential to develop better understanding of the sensitivity in worker performance \cite{yu2015efficient} \cite{kittur2008crowdsourcing}\cite{sorokin2008utility} and task outcome under different competition circumstances, i.e. with respect to different levels of task diversity. To the best of our knowledge, there is no investigation on impact of the different dynamic of tasks supply in the platform as task diversity on tasks success.

The objective of this study is to empirically investigate patterns and impact of task diversity in software crowdsourcing platform in order to improve the success and efficiency of software crowdsourcing. In this study, we first present a motivational example to shed light on the identification of three different patterns of task diversity in software crowdsourcing platform. Then, we propose a conceptual task diversity model, and develop an approach to characterizing and analyzing patterns and effect of task diversity. More specifically, this includes grouping similar tasks, ranking them based on their competition level and identifying the dominant attributes that distinguish among these levels, and then studying the impact of task diversity on task success and worker performance. The empirical study is conducted on more than one year’s real-world data from TopCoder, the leading software crowdsourcing platform with an online community of over 1.5M workers and 55k mini-tasks (website \url{https://www.topcoder.com/}). The evaluation results show that: 1) monetary prize and task complexity are the dominant attributes of task diversity; 2) exists three task diversity patterns (configurations) based on workers behavior: responsive-to-prize, responsive-to-prize-and-complexity and over-responsive-to-prize; 3) responsive-to-prize-and-complexity configuration results in the lowest task failure ratio.

The remainder of this paper is structured as follows. Section II introduces a motivational example that inspires this study. Section III presents related work. Section IV outlines our research design. Section V presents the empirical results. Section VI discusses the key findings of our study. Section VII presents the conclusion and outlines a number of directions for future work.

\section{A Motivational Example}

We suppose that task diversity in the platform is a relevant factor in workers decision making process registrations and submissions for a task. To investigate dynamic patterns of crowdsourced software tasks in terms of task supply and arrival during a specific time-period, a preliminary analysis is conducted using data from Topcoder platform.

\textbf{Dataset.} We use TopCoder dataset from January 2014 to February 2015 which contains 4,907 tasks. We collect tasks' attributes such as dates for registration and submission, task description, technologies and platforms required, and monetary prize. We create attributes like number of days for registration and submission, task complexity proxy by number of words in task description, number of links in task description, binary variables for platforms and technologies, and the number of each of these.

\textbf{Analysis Step.} We group the tasks' attributes per month, creating 14 datasets and for each we perform the following: 1) unsupervised clustering using K-means algorithm where inputs are normalized between 0 and 1. Identify the number of optimal clusters by the ``elbow'' criteria. Each cluster is ranked according to the average number of registrations to identify the level of attraction of workers (competition level). Then, we perform a decision tree classification, where the inputs are tasks' attributes and the label is the ranked cluster. This analysis helps us to identify which are the top task attributes that differentiate among one ranked cluster and another. For all the 14 months, the dominant attributes are prize and task complexity.

\begin{figure*}[hbt!]
\centering
\includegraphics[width=1\textwidth,keepaspectratio]{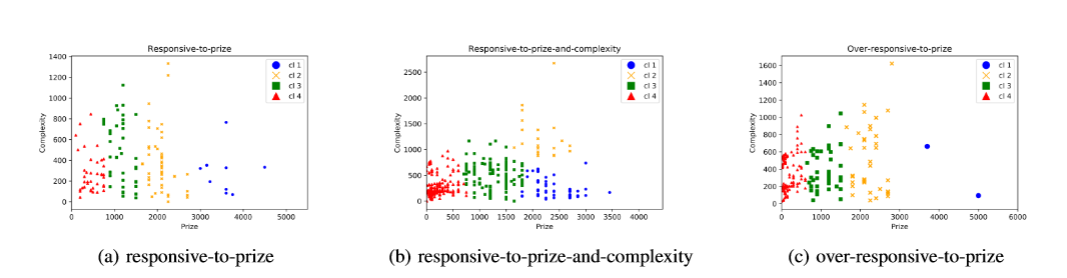}
\caption{Example of responsive-to-prize, responsive-to-prize-and-complexity and over-responsive-to-prize configurations.}
\label{3_patterns}
\end{figure*}

\textbf{Task Diversity Patterns.} We visualize for each month the configuration of clusters with the dominant attributes: prize and complexity, along with their competition levels. The competition level is related to the number of registrations; as more workers register for a task, more competition among them. After analyzing each time-period, we conclude three task diversity patterns (configurations): responsive-to-prize, responsive-to-prize-and-complexity and over-responsive-to-prize. Figure \ref{3_patterns} illustrate the three patterns.

\textit{Responsive-to-prize} in Figure \ref{3_patterns}a, presents the distribution of tasks' clusters for one month, in the x-axis is the task prize and in the y-axis the task complexity, where each color represents a cluster of tasks. On the top right is the legend for the competition level for each cluster, where cl1 groups the cluster with highest competition level (number of registrations), cl2 the second best, cl3 the third and cl4 the fourth. The plot illustrates in blue, the cluster of tasks with higher competition level (number of registrations), 22 in average, followed by 2nd ranked cluster yellow with 21, a 3rd ranked cluster in green, with 16 and the 4th in red with 12. In this time-period, we observe 1) competition level follows a prize order, where we can expect that if a task pays a higher prize it can expect a higher number of registrations; 2) the difference in the average of registrations per cluster does not varies much.

\textit{Responsive-to-prize-and-complexity} in Figure \ref{3_patterns}b, shows the case of a time-period, where not only prize plays a role in the attraction of workers, but also task complexity. Observe that the top ranked cluster, colored in blue, groups tasks with high prize but low complexity, with 18 registrations in average. While the 2nd ranked, colored in yellow, has a high prize and complexity, with 17 registrations. In this time-period we observe that workers attraction follows a task prize, followed by task complexity.

\textit{Over-responsive-to-prize} in Figure \ref{3_patterns}c, illustrates the case of a month where we have outliers of tasks in terms of prize. The top ranked cluster in blue, had an average of registrations of 40, while the 2nd ranked (yellow) just 26, the 3rd ranked (green) 20, and 4th ranked (red) only 8. In this month, we observe that when tasks with high prize, compared to the average prize in that time-period, arrives, it creates a disruption and modifies the behavior of workers. In this case the difference in registrations between the 1st and 2nd ranked clusters is 14 registrations, which is more higher than what we observe in \textit{responsive-to-prize} and \textit{responsive-to-prize-and-complexity} patterns.

After analyzing these patterns, we performed an analysis of the diversity patterns on task success and worker performance, and found that 1) responsive-to-prize configuration provides highest level of task density and workers' reliability in a platform; 2) responsive-to-prize-and-complexity configuration leads to attracting high level of trustworthy workers; 3) over-responsive-to-prize configuration results in highest task stability and the lowest failure ratio in the platform for not high similar tasks.

\section{Related Work}

\subsection{Task Similarity in Crowdsourcing}

Generally, workers tend to optimize their personal utility factor to register for a task \cite{faradani2011s}. It is reported that workers are more interested in working in similar tasks in terms of monetary prize \cite{yang2015award}, context, technology \cite{difallah2016scheduling}, and complexity level. Context switch generates reduction in workers’ efficiency \cite{difallah2016scheduling}. However, workers usually try to register for a greater number of tasks than they can complete \cite{yang2016should}. It is reported that high task similarity level negatively impacts task competition level and team elasticity \cite{saremi2020right}. Combination of these observations lead to receiving task failure due to 1) receiving zero registration for task based on low degree of similar tasks and lack of available skillful worker \cite{yang2015award}, and 2) receiving non-qualified submissions or zero submissions based on lack of time to work on all the registered task by the worker\cite{archak2010money}.

\subsection{Workers Performance in Crowdsourcing}
Software workers’ arrival to the platform and the pattern of taking tasks to completion are factors that shape the worker dynamic in a crowdsourcing platform, however, the reliability in returning the qualified tasks creates the dynamic of the platform. Generally, not only would the award associated with the task influence the workers’ interests in competitions\cite{yin2014monetary}, the number of registrants for the task, the number of submissions by individual workers, and certainly the workers’ historical score rate would directly affect their final performance \cite{lakhani2010topcoder}\cite{saremi2017leveraging}. For newcomers or beginners, there is a time window required to improve and to develop into an active worker \cite{faradani2011s}. Therefore, it is typical that the workers need to communicate with the task owner in order to better understand the problems to be solved \cite{kittur2013future}. Existing studies show that over time, registrants gain more experience, exhibit better performance, and consequently gain higher scores \cite{faradani2011s} \cite{archak2010learning} \cite{kittur2013future}. Still, there are workers who manage not only to win but also to raise their submission-to-win ratio \cite{difallah2016scheduling}. This motivate workers to develop behavioral strategies in Topcoder \cite{archak2010money} \cite{archak2010learning}. Moreover, the ranking mechanism used by TopCoder contributes to the efficiency of online competition and provides more freedom of choice for higher rate workers in terms of controlling competition level \cite{archak2010money}.

\subsection{Modeling Competition in Crowdsourcing}

A few studies have considered game theory in competitive crowdsourcing. For example, Wu et. al. proposed the min-max nature of software competition contributes to the quality and creativity of delivered software \cite{wu2013creative}. Not only the associated monetary prize with the task would influence competition level, but also the number of registrants for the task, number of submissions by individual worker, and of workers historical score rate would directly affect on the final competition\cite{moshfeghi2016identifying}. Such a mutually destructive contest, often called the “Chicken Game” or “Hawk-Dove” \cite{rapoport1966game}, in which less aggressive competitors (chicken or dove) will yield to aggressive competitors. In such crowdsourced environments, task duration can be used to identify careless workers \cite{moshfeghi2016identifying}.

Moreover, according to competitive exclusion principle\cite{hardin1960competitive}, sometimes referred as Gause's law \cite{pocheville2015ecological}, in competing for limited resources, competitors with the slightest advantage over others frequently win and eventually dominate in the system.

In a CSD platform, a competitive environment not only influences the decisions of contestants regarding which tasks to register and submit but also how these react to their peers. Hu and Wu modeled single-round matches through a game theory framework where participants find bugs in a program \cite{hu2015game}. The authors found that the major factors for contestants’ decisions are effort cost and probability of making a successful challenge, reinforcing the finding in \cite{hu2014game}, and that contestants with a high rating are more likely to launch challenges against lower ones.

\section{Research Design}
Driven by task diversity patterns introduced in Section II, this study aims at further investigating the dominant influencing attributes and effect of task diversity patterns on crowdsourcing task success. This section presents a conceptual task diversity model, and an approach to characterizing and analyzing patterns and effect of task diversity.  

\subsection{Conceptual Task Diversity Model}

\begin{figure}[hbt!]
\centering
\includegraphics[width=0.72\columnwidth,keepaspectratio]{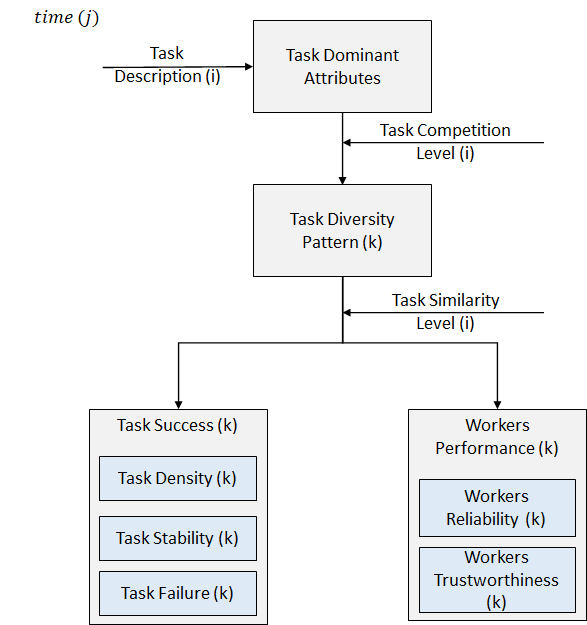}
\caption{Conceptual Task Diversity Model}
\label{tam}
\end{figure}

To develop better understanding on the dynamic patterns in task supply and execution, we formulate a conceptual model to represent the simplified underlying relationship among various dynamic attributes in software crowdsourcing processes. It is based on the Award-Worker behavior model in \cite{yang2015award}, which models the factors that influence the behavior of workers to register and submit a task. In this study, we extend the Award-Worker behavior model to represent and investigate the different task diversity patterns in a specific time-period, in order to further investigate empirical relationships among them in quantitative manners. The proposed conceptual model is illustrated in Figure \ref{tam}. Next, we introduce a few working definitions used in this model.

\subsubsection{Task Attributes}

Task attributes are the subset of task dynamic attributes that influences task diversity pattern. The list of task attributes applied in this study is summarized in Table \ref{metrics}.

\subsubsection{Task Competition Level}

Task competition level is defined as the relative level of competition for a task in terms of attracting crowd workers \cite{saremi2017leveraging}\cite{yang2015award}\cite{archak2010money}, and is measured using the number of registrations on that task. Intuitively, larger number of registrations imply higher level of competition on the corresponding task. 

\subsubsection{Task Diversity Pattern} Task diversity pattern is defined as the configuration of tasks supply in a time-period with respect to different competition levels in a CSD platform. 

\subsubsection{Task Similarity Level} Task similarity level is defined as the degree of task similarity between a set of simultaneously open tasks. To analyze task similarity, we calculate the tasks’ local distance from each other to analyzed task similarity factor, based on textual task requirements. More specifically, the calculation follows the following detailed definitions.

\textit{Def. 1}: Task local distance ($ {D_{i}S{j}} $), is a tuple of all tasks’ attributes in the dataset. In respect to introduced variables in Table \ref{metrics}, task local distance is defined as:

\[
\begin{split}
{D_{i}S_{j}} = {(RD, RTSD, D, P, TC,}\\
{\#Tech, PLT, \#PLT, \#Tech, TDI)}
\end{split}
\]

\textit{Def. 2}: Task Similarity Factor $ {TS_{ij}} $ is dot product and magnitude of local distance of two tasks:

\[
{TS_{ij}} = \frac{\sum_{i,j=0}^{n}{D_{i}S_{j}(T_{j}, T_{i})}} 
{\sum_{i=0}^{n} \sqrt(D_{i}S_{j}(T_{i}) * \sum_{j=0}^{n} \sqrt(D_{i}S_{j}(T_{j})}
\]

\subsubsection{Task Success} In this study, task success is defined as a tuple of task density ($ {TD_{i}} $), task stability ($ {TSL_{i}} $), and task failure ($ {TF_{i}} $). Task density is a measure to determine the probability of attracting workers in a dynamic competitive market base on resource share in the market, while task stability is the probability of receiving a submission by attracted workers. To measure the impact of task diversity pattern on task success, we consider task similarity factor among tasks in same configuration to compare performance metrics among similar tasks.
      
\[
{Task Success_{i}}  = {({TD_{i}},{{TSL_{i}}, {TF_{i}}})}
\]

for a given time-period $j$, similarity level $i$ and task diversity pattern $k$.

Task density, task stability, and task failure are defined as below:

\textit{Def. 1}: Task Density $ {TD_{j}} $ is the ratio of similar arrival tasks per time-period $ {AT_{i}} $ in the platform by total number of arrival tasks $ {CAT_{i}} $ per task configuration in the platform,

\[
{TD_{j}} = \frac{\sum_{i=0}^{n}{AT_{i}}} 
{\sum_{i=0}^{n}{CAT_{i}}}
\]


Average submissions for task $i$ in the same group of similar tasks per task configuration will illustrates task stability to be completed in the platform. 

\textit{Def. 2}: Task Stability Level, ${TSL_{j}}$, measures average submissions of arrival similar tasks per task configuration.

\[
{TSL_{j}} = \frac{\sum_{i=0}^{n}{S_{i}}} 
{\sum_{i=0}^{n}{CAT_{i}}} 
\]


Average failed similar tasks per task configuration in the same month represents task failure ratio.

\textit{Def. 3}: Task Failure, ${TF_{j}}$, per similarity level per task configuration is number of failed tasks (${ft_{i}}$) in the platform, which arrived at the same calendar month per configuration, ${Cf_{i}}$ :

\[
{TF_{j}} = \frac{\sum_{i=0}^{n}{ft_{i}}} 
{\sum_{i=0}^{n}{Cf_{i}}} 
\]

 
\subsubsection{Worker Performance} Worker performance is defined as tuple of  workers' reliability in making a submission when registered for a task ($ {RL_{k}} $) and  workers' trustworthiness in making a qualified submission for the registered task ($ {TL_{k}} $). Note that each worker is a tuple of number of registration ($ {WR_{k}} $), number of submissions ($ {WS_{K}} $), and number of valid submission ($ {WVS_{k.}} $).

Similarly, to measure the impact of task diversity pattern on worker performance, we consider task similarity factor among tasks in same configuration to compare performance metrics among similar tasks. 

\[
{Worker Performance_{k}}  = {({RL_{k}},{{TL_{k}}})}
\]

for a given time-period $j$, similarity level $i$ and task diversity pattern $k$.

Worker reliability and worker trustworthiness are defined as follow:

\textit{Def. 6}: Workers’ Reliability Level, ${RL_{k}}$, measures average reliability for worker $k$ to register for similar tasks in the same task configuration, ${CR_{i}}$. 

\[
{RL_{k}} = \frac{\sum_{i=0}^{n}{S_{i}}} 
{\sum_{i=0}^{n}{CR_{i}}}
\]


Tasks owners trust on receiving qualified submissions after receiving a registration by a reliable worker. This is defined as the ratio of receiving a valid submission by workers per similarity level in a task configuration to make a submission for the task.

\textit{Def. 7}: Workers’ Trust Level, ${TL_{k}}$, measures average valid submission, ${VS_{i}} $, for worker $k$ to register for similar tasks in the same task configuration, $ {CR_{i}} $.

\[
{TL_{k}} = \frac{\sum_{i=0}^{n}{VS_{i}}} 
{\sum_{i=0}^{n}{CR_{i}}} 
\]


\subsection{Research Questions}
Three research questions are formulated as following:

\begin{itemize}

    \item \textit{RQ1 (Task Diversity Patterns)}: How do different task diversity patterns (configurations) distribute in a competitive CSD?
    
     This research question aims at providing general overview of task diversity based on the dominant tasks attributes and competition levels per time-period (monthly) in the CSD platform;
    
    \item \textit{RQ2 (Task Success)}: How does different task diversity patterns impact tasks success?
    
    Understanding task density, task stability and task failure ratio per group of task type can be good measure to indicate tasks success;
    
    \item \textit{RQ3 (Worker Performance)}: How does different task diversity patterns impact workers performance?
    
    The ratio of attracting reliable workers to compete on the tasks per task type and trust-able to return valid submissions by the workers represent workers’ consistency to work on tasks.

\end{itemize}

\subsection{Dataset}

The dataset from TopCoder contains 403 individual projects including 4,907 component development tasks (ended up with 4,770 after removing tasks with incomplete information) and 8,108 workers from January 2014 to February 2015 (14 months). Tasks are uploaded as competitions in the platform, where Crowd software workers would register and complete the challenges. On average, most of the tasks have a life cycle of one and half months from first day of registration to the submission’s deadline. When the workers submit the final files, it will be reviewed by experts to check the results and grant the scores.

The dataset contains tasks attributes such as technology, platform, task description, monetary prize, days to submit, registration date, submission date, and the time-period (month) on which the task was launched in the platform. In this step, task's (workers) performance metrics are not included. Then, we create attributes, such as days for registration (RD), days from registration to submission (RTSD), task complexity (TC), which is proxy by the number of words in the task description, number of links in task description (\#L), number of platforms (\#PLT), number of technologies (\#Tech) and the creation of binary variables for each technology (Tech) and platform (PLT) required in each task, where:

\[
f(x,s) =
	\begin{cases}
        1 &\parbox[t]{.35\textwidth}{task x requires technology or platform s} \\
        0 & \text{otherwise}
    \end{cases}
\]

We also created a task difficulty index (TDI) that incorporates the number of words of the task description, the automated readability score of the task description and the number of technologies required in task. Each component of this index as it has a higher value, may represent more difficulty. Each component was normalized between 0 and 1 and then we calculate the mean of three components per task.

We create a dataset of task attributes for each time-period (month). The tasks attributes used in the analysis are presented in top section of Table \ref{metrics}.
 
\begin{table*}[!ht]
\caption{Summary of Metrics Definition} 
\centering 
\begin{tabular}{p{2cm} p{4cm} p{10cm}}
\hline
Type & Metrics & Definition \\ 
\hline
\multirow{13}{*}{Tasks attributes} 
& Days for Registration (RD) & Number of days to for the task. \\ 
& Days from Reg. to Sub.(RTSD) & Number of days to submit the task once registration is closed.  \\ 
& Duration (D) & Total available days from registration date to submissions deadline. \\ 
& Monetary Prize (P) & Monetary prize (Dollars) in task description. \\ 
& Task Complexity (TC) & Number of words in task description.  \\ 
& Link count (\#L) & Number of links in task description.  \\ 
& Technology (Tech) & Required programming language to perform the task.  \\ 
& Platform (PLT) & Associate platform that a task is performing. \\
& Platforms (\#PLT) & Number of platforms used in task. \\ 
& Technologies (\#Tech) & Number of technologies used in task.  \\ 
& Task Difficulty Index (TDI) & Index that incorporates the number of words and readability score in the task description, and the number of technologies required to perform the task. \\
\hline
\multirow{7}{*}{Tasks outcome} 
& Task Status & Completed or failed task. \\
& \# Registration (R)  & Number of registrants that are willing to compete on total number of tasks in specific period of time.  \\ 
& \# Submissions (S) & Number of submissions that a task receives by its submission deadline in specific period of time. \\ 
& \# Valid Submissions (VS) & Number of submissions that a task receives by its submission deadline and passed the peer review in specific period of time. \\ 
\hline
\label{metrics}
\end{tabular}
\end{table*}

\subsection{Empirical Study Design}

\begin{figure}[ht!]
\centering
\includegraphics[width=0.72\columnwidth,keepaspectratio]{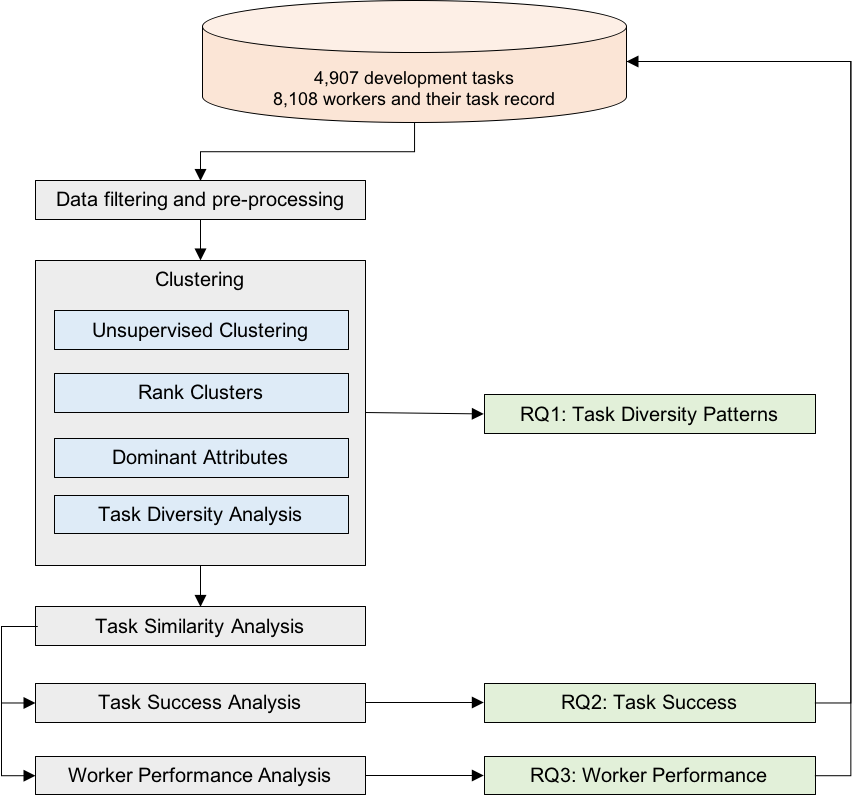}
\caption{Main flow of the proposed framework and relationship to research questions.}
\label{diagram}
\end{figure}

In this work we followed a similar task content similarity clustering approach as \cite{fu2017competition}, which includes task features such as task duration, award, technology and platform and applied a K-Means algorithm. As Figure \ref{diagram} illustrates, our approach contains the following steps.

\subsubsection{Clustering Analysis}

We perform a clustering analysis to identify the configurations of groups of tasks for each time-period (month). For each dataset we perform the following: first, an unsupervised clustering for each time-period where each task is a vector of attributes and these were normalized transforming the original values to values between 0 and 1. The clustering method used is K-Means. To identify the number of optimal clusters we use the ``elbow'' criteria. This method observes the variance explained as the number of clusters increases. We select the ``optimal'' number of clusters, when the marginal gain of adding one more cluster drops, creating an ``elbow'' in the plot that illustrates the inverse of variance explained per number of clusters. The number of optimal clusters for each month varies.

Second, we rank each cluster by the competition level, which is the mean of registrants per cluster. In this step, we have every observation (task) assigned to a cluster where cluster 1 is the top ranked, cluster 2 the second best ranked, successively.

Third, we identify the dominant attributes that differentiate ranked clusters to one another. We performed a decision tree classification where the input are the attributes created and the classification attribute is the ranked cluster from previous step. The dataset is divided in two samples: train (75\%) and test (25\%) and different decision trees were run, looking for the least number of attributes that generate the highest accuracy in the classification. For each decision tree, we select the attributes with higher importance as the dominant ones. After performing these three steps for each month dataset, we identified that the frequent dominant task attributes that differentiate among the ranked clusters are prize and complexity.

\subsubsection{Task Diversity Analysis}

We analyze each time-period by visualizing the dominant attributes identified previously and included the ranked cluster label of each observation for each time-period. In our analysis the frequent dominant attributes for each time-period were monetary prize (x-axis) and complexity (y-axis). Each observation is colored based on the cluster's rank. For example, tasks in cluster 1 (top ranked) are colored in blue, cluster 2 (2nd best) in yellow, cluster 3 in green and cluster 4 in red. This allows to visualize the distribution of clusters considering the dominant attributes (prize and complexity) and how well the clusters performed against each other (competition level).

\subsubsection{Task Success Analysis}

We study the impact of the different task diversity patterns (configurations) on tasks success by looking into task density, task stability and task failure ratio. To do so, the number of similar arrival task as well as similar open task per task configuration was analyzed. Then, task density, task stability and task failure per task similarity level and task configuration was analyzed.

\subsubsection{Worker Performance Analysis}

We investigate the impact of the different task diversity patterns (configurations) on workers performance by looking into workers' reliability and trust. The probability of a worker make a submission after registered for a task will be reported as worker reliability, and the probability of a submission passes the peer review and labeled as valid submission provides workers trust. In TopCoder, crowd worker’s reliability of competing on the tasks is measured based on last 15 competitions workers registered and submitted. For example, if a worker submitted 14 tasks out of 15 last registered tasks, his reliability is 93\% (14/15). 

This analysis is relevant since each task configuration may impact the submissions of workers, for example, in the pattern over-reactive-to-prize we observed that tasks with high prize above the average, attract more competition level, however how many of these registrants will actually submit their task.

\section{Empirical Results}

\subsubsection{Task diversity patterns (RQ1)}

We perform a longitudinal analysis of task diversity patterns analyzing 14 months, January 2014 to February 2015.

The task attributes importance are summarized on Table \ref{months_top_attributes} for each time-period, the rest of attributes not shown had an importance weight zero. As it is observed, the frequent most dominant is monetary prize, followed by complexity. The third dominant factor was difficulty index, however this attribute was only relevant in January and December 2014. While prize and complexity were more frequent. These attributes are key to differentiate among different competition levels and used to identify task diversity patterns (configurations).

\begin{table*}[!ht]
\begin{threeparttable}
\caption{Task attributes importance} 
\centering 
\begin{tabular}{p{0.98cm}p{0.78cm}p{0.78cm}p{0.78cm}p{0.78cm}p{0.78cm}p{0.78cm}p{0.78cm}p{0.78cm}p{0.78cm}p{0.78cm}p{0.78cm}p{0.78cm}p{0.78cm}p{0.78cm}}
\hline
Attribute & Jan'14 & Feb'14 & Mar'14 & Apr'14 & May'14 & Jun'14 & Jul'14 & Aug'14 & Sep'14 & Oct'14 & Nov'14 & Dec'14 & Jan'15 & Feb'15\\
\hline
Prize &0.381&0.952&0.785&0.994&0.962&1&0.964&0.976&0.971&0.886&0.988&0.83&1&0.486\\
Complexity &0.401&0.048&0.172&&0.013&&0.025&0.024&0.015&0.101&&0.053&&0.222\\
Difficulty &0.218&&0.013&&&&0.006&&&&&0.102&&\\
\#L&&&0.003&&&&&&&&0.004&&&\\
\#PLT&&&0.013&&&&&&&&0.008&0.009&&\\
RD&&&&&0.019&&0.005&&0.015&&&&&\\
RTSD&&&&&&&&&&0.005&&0.005&&\\
\hline
Express&&&0.014&&&&&&&&&&&\\
Web Services&&&&0.006&&&&&&&&&&\\
.NET&&&&&0.007&&&&&&&&&\\
Java&&&&&&&&&&0.008&&&&\\
JavaScript&&&&&&&&&&&&&&0.292\\
\hline
\label{months_top_attributes}
\end{tabular}
\begin{tablenotes}
      \small
      \item \#L = number of links in task description, \#PLT = number of platforms used in tasks, RD = Days for Registration, RTSD = Days from Registration to Submission; Express, Web Services, .NET, Java and JavaScript are technologies required in task.
\end{tablenotes}
\end{threeparttable}
\end{table*}

After analyzing the plots for all time-periods, we found three task diversity patterns (configurations): responsive-to-prize, responsive-to-prize-and-complexity and over-responsive-to-prize as shown in Figure \ref{3_patterns}. These names were given based on the response of workers in each configuration. The task diversity patterns (configurations) were defined as follows:

\begin{itemize}
    \item \textit{Responsive-to-prize} (RP): time-period where competition level follows prize order. In this configuration we expect that tasks with higher monetary prize attract higher competition level, while the average competition level per task cluster does not vary much;

    \item \textit{Responsive-to-prize-and-complexity} (RPC): time-period where competition level follows both prize and task complexity order. In this configuration we expect that tasks attract competition level not only based on monetary prize but also how complex a task is;

    \item \textit{Over-responsive-to-prize} (ORP): time-period where there are few tasks with monetary prize above the average, as an outlier, which are attracting the highest competition level. In this configuration a considerable difference in task competition level between the 1st and 2nd ranked clusters is expected.
\end{itemize}

Table \ref{months_labels} summarizes the tasks' configurations for each month. As it can be observed, 5 months had a RP configuration, 3 were RPC and 6 were ORP. 

\begin{table*}[!ht]
 \begin{threeparttable}
\caption{Task diversity pattern per month.} 
\centering 
\begin{tabular}{p{0.98cm}p{0.78cm}p{0.78cm}p{0.78cm}p{0.78cm}p{0.78cm}p{0.78cm}p{0.78cm}p{0.78cm}p{0.78cm}p{0.78cm}p{0.78cm}p{0.78cm}p{0.78cm}p{0.78cm}}
\hline
Conf. & Jan'14 & Feb'14 & Mar'14 & Apr'14 & May'14 & Jun'14 & Jul'14 & Aug'14 & Sep'14 & Oct'14 & Nov'14 & Dec'14 & Jan'15 & Feb'15\\
\hline
RP & & \checkmark &  \checkmark &  \checkmark &  \checkmark & &  \checkmark & & & & & & &\\
RPC &  \checkmark & & & & & & & & & & &  \checkmark & & \checkmark \\
ORP & & & & & &  \checkmark & &\checkmark &  \checkmark &  \checkmark &  \checkmark & &  \checkmark &\\
\hline
\label{months_labels}
\end{tabular}
\begin{tablenotes}
      \small
      \item RP = responsive-to-prize, RPC = responsive-to-prize-and-complexity and ORP = over-responsive-to-prize.
\end{tablenotes}
\end{threeparttable}
\end{table*}

In the 1st half of 2014, we had RP months, which means that we observed an expected behavior: higher attraction of workers (competition level) as the prize is higher.

In the 2nd half of the 2014, we observed ORP pattern, where we had tasks with high prize, above the average, creating a disruption and attracting an unusual number of workers to register to these.

The RPC months were less frequent. These periods, the workers decision to register is not only determined by the prize, but also the complexity of the tasks. In this case, a task manager should consider this dynamic if they want to have success when launching high-complex tasks given this dynamic of tasks supply.

\textit{Finding 1.1}: The task diversity patterns identified were responsive-to-prize (RP), which was present in the 1st half of 2014, over-responsive-to-prize (ORP) frequent in the 2nd half of 2014, while responsive-to-prize-and-complexity (RPC) was the least frequent, found only in 3 months.

\subsubsection{Task Success (RQ2)}

In order to have a better understanding of tasks success, we studied distribution of task density in the different tasks similarity levels per task diversity pattern (configuration). By considering the task similarity among tasks, we can compare similar tasks under different task configurations. To do so, we created three different datasets based on the identified task diversity pattern, then we calculated the task similarity among the available task per diversity pattern. Since it is reported that, the task similarity over 60\% impacts on task performance as success \cite{saremi2020right}, we analyzed defined metrics in section IV-A for task similarity degree of 60\%, 70\%, 80\% and 90\% per month. Table \ref{taskdistribution} summarises the distribution of task per task configuration. Table \ref{taskdistribution} reports the unique number of each task pair per similarity level per month (i.e. $ {TS_{ij}} = {TS_{ji}} $) .

\begin{table}[!ht]
\caption{Summary of tasks distribution per diversity pattern} 
\centering 
\begin{tabular}{p{4cm}p{1cm}p{1cm}p{1cm}}
\hline
Diversity Pattern & RP & ORP & RPC \\
\hline

\# Tasks & 1901 & 2383 &  486  \\

Average \# tasks per month & 292 & 315 & 162  \\

Std of tasks per month & 90 & 16 &  50  \\

\# 60\% Task pair similarity & 366 & 546 & 180  \\

\# 70\% Task pair similarity & 600 & 189 & 146  \\

\# 80\% Task pair similarity & 753 & 310 & 122  \\

\# 90\% Task pair similarity & 827 & 760 & 90  \\
\hline
\label{taskdistribution}
\end{tabular}
\end{table}

\begin{figure}[ht]
\centering
\includegraphics[width=0.9\columnwidth,keepaspectratio]{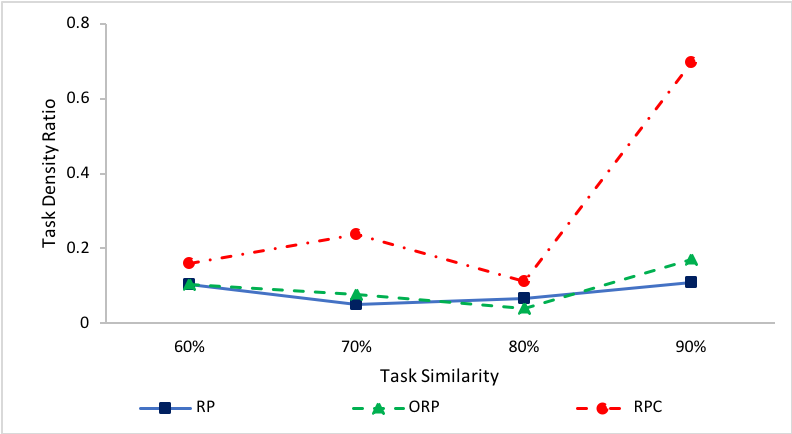}
\caption{Average Task Density per Task Similarity and Diversity Configuration.}
\label{Density}
\end{figure}

According to Figure \ref{Density}, all three patterns correspond to an increasing trend of task density as task similarity increases after 80\%. While RP and ORP configurations seem to follow a similar shape, RPC configuration started with highest task density among tasks with 60\% similarity with 28\% task density. This pattern raised, containing the highest tasks density among all three configurations for all levels of similarity shown, it reached 69\% of task density when tasks with 90\% similarity arrive in the platform. RP configuration follows a steady increase in providing task density with increasing task similarity in among open tasks in the platform. While it provides around 4\% and 6\% density for task similarity of 70\% and 80\% respectively, the density level increased to 10\% for tasks with similarity level of 90\%. On the other hand, tasks density dropped for the ORP configuration when task similarity increased from 60\% to 80\% from 10\% to 3\%, and increased to 17\% for tasks with 90\% similarity. We ran repeated measure one-way ANOVA test on the task density result from the three different task configurations. Based on ANOVA test result, the task density is not significantly different across the three task diversity pattern since the p-value is 0.14.

\begin{figure}[ht!]
\centering
\includegraphics[width=0.9\columnwidth,keepaspectratio]{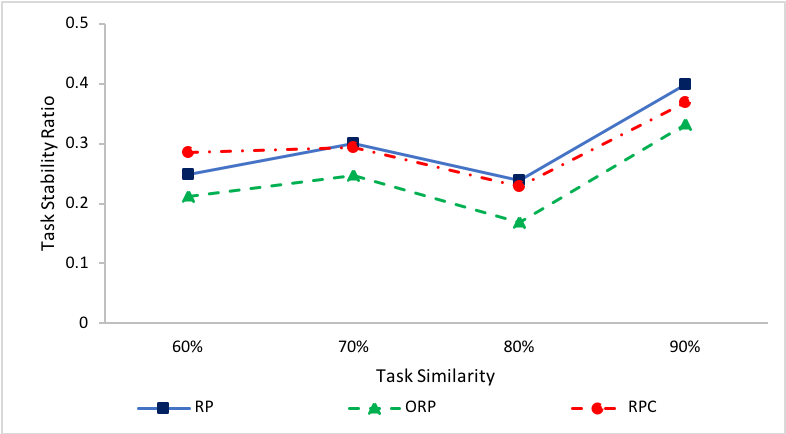}
\caption{Average Task Stability per Task Similarity and Diversity Configuration.}
\label{Stability}
\end{figure}

Another factor that impacts on task robustness is task stability. We investigate the level of task stability per similarity level per task configuration. As Figure \ref{Stability} shows, ORP configuration is providing lowest task stability in 80\% similarity with almost 16\% stability, while RP and RPC have similar stability levels through all the levels of task similarity. Interestingly all tasks configurations are providing between 20\% and 30\% of task stability when we have a task arrival with 80\% similarity or less. While task stability increased for the three configurations for 90\% similarity, where RP and RPC reached 39\% and 36\%, respectively, and ORP 33\%. All three configurations follow the same pattern through the different levels of task similarity. According to the one-way ANOVA test on the task stability result from the three different task configurations, the task stability is not significantly different across the three task diversity pattern with p-value of 0.43.

\begin{figure}[ht!]
\centering
\includegraphics[width=0.9\columnwidth,keepaspectratio]{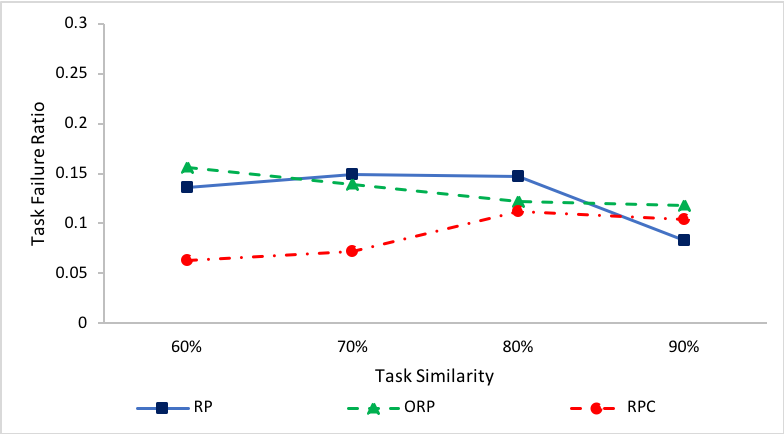}
\caption{Average Task Failure per Task Similarity and Diversity Configuration.}
\label{Failure}
\end{figure}

After understanding different patterns of task density and task stability for each configuration, we investigate the task failure ratio per task similarity level per configuration. Figure \ref{Failure} illustrates the distribution of task failure ratio per similarity level per configuration. Interestingly, RPC provides on average lower task failure ratio in the platform. Tasks with similarity level of 60\% face almost 6\% failure under RPC configuration. It increased to 7\% failure ratio for tasks with 70\% similarity and 11\% for similarity of 80\%, and dropped to 10\% with 90\% similarity. While RP configuration failure ratio for 60\% and 80\% task similarity is very similar and it varies from 13\% to 14\%. It drops to 8\% tasks with similarity of 90\%. ORP configuration started with 15\% task failure for 60\% task similarity, and continue a decreasing pattern passing to 13\% task failure for tasks with 70\% similarity, and 12\% for tasks with 80\%, and 11\% for 90\% of similarity. The result of the ANOVA test showed that task failure ratio is significantly different across all three task configurations
(i.e. p-value is 0.05).

\textit{Finding 2.1}: For tasks with similarity level between 60\% and 80\%, responsive-to-prize-and-complexity (RPC) configuration provides the lowest failure ratio; while responsive-to-prize (RP) configuration has lowest failure with 80\% to 90\% task similarity.

\subsubsection{Worker Performance (RQ3)}

\begin{figure}[ht!]
\centering
\includegraphics[width=0.9\columnwidth,keepaspectratio]{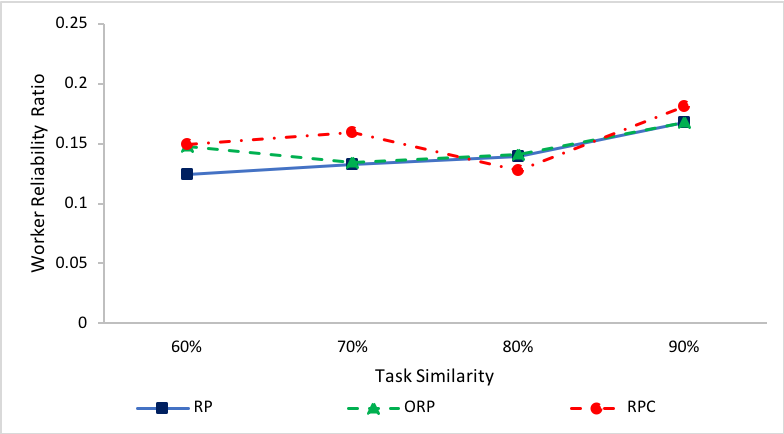}
\caption{Average Workers' Reliability per Task Similarity and Diversity Configuration.}
\label{reliability}
\end{figure}

Workers’ performance can be analyzed based on workers’ reliability and trust in returning a valid submission. Therefore, we analyzed attracting reliable workers and receiving valid submissions per task based on the open tasks in different task configuration. Figure \ref{reliability} illustrates the average workers’ reliability per task configuration. As it is shown in Figure \ref{reliability}, RPC configuration attracts the most reliable workers for 60\% to 70\% and 90\% similar tasks. While, RP and ORP attracts reliable workers at from 70\% to 80\% task similarity. Task in RP and ORP configurations attracted workers with 12\% and 14\% reliability respectively for 60\% task similarity. Both configurations attract workers with 14\% reliability for tasks with 70\% and 80\% similarity and increase to 16\% in 90\% similarity. ANOVA test showed that worker reliability is not significantly different among all three task configurations (i.e. p-value is 0.62).

\begin{figure}[ht!]
\centering
\includegraphics[width=0.9\columnwidth,keepaspectratio]{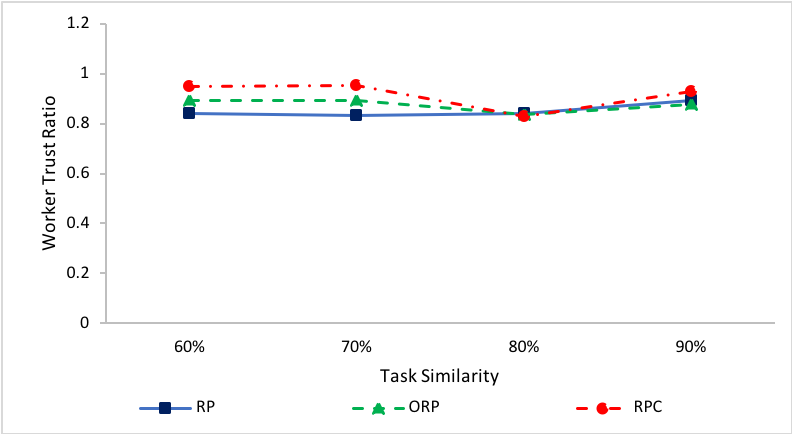}
\caption{Average Workers' Trust per Task Similarity and Diversity Configuration.}
\label{trust}
\end{figure}

Besides attracting reliable workers to make a submission, it is important to trust on the workers’ submissions. Hence, we investigate workers’ trust ratio in returning valid submission. Figure \ref{trust} represents the distribution of workers’ trust ratio among different task similarity per configuration. Among tasks with similarity of 60\% and 70\%, RPC configuration received the highest trust ratio of 94\% and 95\%, followed by ORP with 89\% for  60\% and 70\% task similarity, and RPC with 84\% and 83\% of trust. Interestingly in task similarity of 80\%, workers’ trust in three configurations drop to a similar level between 82\% and 84\%. And finally, for task similarity of 90\%, task trust increased to 89\%, 87\%, 92\% for RP, ORP and RPC, respectively. Interestingly while RP configuration is following a linear pattern among different task configurations, ORP and RPC have a U-shape pattern. ANOVA test showed that different task configurations are not significantly influencing workers’ trust (i.e. p-value is 0.14).

\section{Discussion}

\subsection{Task Diversity Patterns}

The dominant task attributes that differentiate among one competition level (number of registrations) and another, are monetary prize and complexity, reinforcing findings on previous research \cite{lakhani2010topcoder, thuan2016factors, kittur2013future, faradani2011s, archak2010money}.

These dominant attributes were input for the detection of task diversity patterns. We identified three configurations: RP, RPC and ORP. During January 2014 and February 2015, we identified that the 1st half of 2014 was frequent a RP configuration, while in the 2nd half presented an ORP pattern. Only in three months was observed a RPC, finding 1.1.

\subsection{Task Success}

To successfully crowdsource a software project in a CSD platform, not only, it is important to fully understand task density in the platform at any point of time. But also, it is vital to know the task stability and failure ratio in the platform. This research investigated these factors based on task similarity level per identified task configuration. It is reported that task similarity impacts highly on task competition level, task stability level and consequently project success \cite{saremi2020right}. According to the statistical analysis, we could not find a significant different among the task diversity patterns when analyze task density and stability. However, as it is reported in finding 2.1, for task with similarity between 60\% and 80\%, competing in the responsive-to-prize-and-complexity (RPC) configuration is the best strategy in terms of task failure.

\subsection{Worker Performance} 

To assure of having a successful project in CSD, beside the importance of attracting reliable workers whom make a submission, it is crucial to make sure to at least attract one trustworthy worker who not only makes a submission but also makes a valid submission. The result of investigating workers performance under different task diversity patterns (configurations) presented that there is no statistical difference among the task diversity patterns when analyze workers' trust and reliability.

\subsection{Threats to Validity}

First, the study only focuses on competitive CSD tasks on the TopCoder platform. Many more platforms do exist, and even though the results achieved are based on a comprehensive set of about 5,000 development tasks, the results cannot be claimed externally valid. There is no guarantee the same results would remain exactly the same in other CSD platforms.

Second, there are many different factors that may influence tasks similarity, task diversity patterns and workers’ decision in task selection and completion. Our similarity algorithm and task diversity patterns approach are based on known task attributes in TopCoder. Different similarity algorithm and task diversity patterns approaches may lead us to different but almost similar results.

Third, the result is based on tasks only. Workers network and communication was not considered in this research. In future we need to add this level of research to the existing one.

\section{Conclusion and future work}

To understand the probability of a tasks success in a crowdsource platform, not only one should understand the task diversity patterns but also, they need to understand the tasks success per point of time and workers performance in taking a task and returning a valid submission. This research investigated task diversity by applying clustering method based on the dominant task attributes: prize and complexity and observed the competition level along these attributes. Then analyzed both tasks success and workers performance per configuration based on task similarity level.

Based on statistical analysis, this study can only support that responsive-to-prize-and-complexity configuration results in the lowest task failure ratio for tasks with similarity level between 60\% and 80\%.

In future, we would like to evaluate our finding in crowdsourced software development practice and testing the scalability of them in real time.

\bibliography{mybibfile.bib}

\end{document}